\documentstyle[aps,epsfig,multicol]{revtex}
\def\beginwide{
        \end{multicols} \vspace*{-0.5cm} \noindent
        \rule{3.5in}{.1mm}\rule{.1mm}{5mm} \widetext \medskip }
\def\beginwidetop{
        \end{multicols} \vspace*{-0.5cm} \noindent
        \widetext \medskip }
\def\endwide{
        \hspace*{3.35in}~\rule[-5mm]{.1mm}{5mm}\rule{3.5in}{.1mm}
        \begin{multicols}{2} \vspace*{-1.0cm} \noindent }
\def\endwidebottom{
        \begin{multicols}{2} \vspace*{-1.0cm} \noindent }

\begin{document}

\draft

\title{Boundary effects on flux penetration in disordered superconductors} 

\author{Andr\'e A. Moreira$^{1}$, 
Jos\'e S. Andrade Jr.$^{1}$, Josu\'e Mendes Filho$^1$,
and Stefano Zapperi$^{2}$}
\address{$^1$Departamento de F\'{\i}sica, Universidade Federal do
Cear\'a, 60451-970 Fortaleza, Cear\'a, Brazil.\\
$^2$INFM UdR Roma 1 and SMC, Dipartimento di Fisica, Universit\`a 
``La Sapienza'', P.le A. Moro 2, 00185 Roma, Italy.
}

\date{\today}

\maketitle
 
\begin{abstract}
We investigate flux penetration in a disordered type II superconductor by
molecular dynamics simulations of interacting vortices. We focus on the
effect of different boundary conditions on the scaling laws for flux front
propagation. The numerical results can be interpreted using a coarse grained
description of the system in terms of a non-linear diffusion equation. We
propose a phenomenological equation for the front position that captures the
essential behavior of the system and recovers the scaling exponents.
\end{abstract}

\pacs{PACS numbers: 74.60.Ge, 05.45.-a, 47.55.Mh}

\begin{multicols}{2}

\section{Introduction}

In recent years, the discovery of high temperature superconductors has
generated a renewed interest in understanding the magnetization properties of
type II superconductors \cite{Blatter}. The magnetization process is usually
described in terms of the Bean model \cite{bean}: as the magnetic flux enters
into the sample from the boundaries, quenched disorder is responsible for the 
formation of a constant flux gradient. 
The Bean model provides a phenomenological picture of
average magnetization properties, such as hysteresis and thermal
relaxation \cite{kim}, but does not describe local space-time
fluctuations.  Indeed, recent experiments showed that these fluctuations are
not only common, but also large, spanning several length scales: the flux line
dynamics is intermittent, taking place in avalanches \cite{avalanches}, and
flux fronts are fractal \cite{rough,fractal,fractal2}.

A widely used modeling strategy to describe the fluctuations in the
magnetization process consists in molecular dynamics (MD) 
simulations of interacting
vortices, pinned by quenched random impurities
\cite{Brass,nori2,pla,nori4}. With this approach it has been
possible to model flux profiles \cite{nori2}, hysteresis \cite{nori2},
avalanches \cite{pla,nori4} and plastic flow \cite{Brass,nori4}. One of
the aims of these studies \cite{nori2} is to link
the macroscopic behavior, as described for
instance by generalized Bean models, to the microscopic vortex dynamics.

Recently, we have shown that the flux penetration due to interacting vortices
in a disordered superconductor can be described by a disordered non-linear
diffusion equation \cite{zms}. The equation can be obtained performing a
coarse-graining of the microscopic equation of motion of the vortices.  In
the absence of pinning, the equation reduces to the model introduced in
Ref.~\cite{dorog}. This model has been solved  analytically  to provide
expressions for the dynamics of the front for different boundary 
conditions\cite{dorog,gilc}.
When quenched disorder is included in the diffusion equations, flux fronts
are pinned in agreement with MD simulations. Varying the parameters of the
equation, we observe a crossover from flat to fractal flux fronts, consistent
with experimental observations. The value of the fractal dimension suggests
that the strong disorder limit is described by percolation \cite{zms}.  
In the weak disorder limit, we recover the analytical results derived in
Refs.~\cite{dorog,gilc}. Using this description, we can thus directly link
continuum theories, for which analytical solutions are possible, to the
microscopic equations used in MD simulations \cite{zms}.

In this paper we systematically analyze the effect of different boundary
conditions on the propagation of the flux front. A similar study was
presented in Refs.~\cite{dorog,gilc} in the framework of non-linear flux
diffusion in the absence of disorder. We show here that the results are in
agreement with MD simulations for various different boundary
conditions. Next, we analyze the effect of disorder by varying the pinning
strength in the MD system. The results are then interpreted
theoretically by means of the non-linear diffusion equation
\cite{zms,dorog,gilc}.  Finally, we propose a phenomenological equation for
the front position that is able to capture in a simple way the behavior of
the system, recovering the numerical results for the different boundary
conditions.

\section{Vortex dynamics model}

In an infinitely long cylinder, flux lines can be represented as a set of
interacting particles performing an overdamped motion in a random pinning
landscape \cite{Brass,nori2,pla,nori4}. The equation of motion
for each flux line $i$ is given by
\begin{equation}
\Gamma \vec{v}_i = \sum_j \vec{J}(\vec{r}_i - \vec{r}_j)+
\sum_p \vec{G}[(\vec{R}_p-\vec{r}_i)/l]~,
\label{eq:vf}
\end{equation}
where the effective viscosity is obtained from material parameters as
$\Gamma=\Phi_0 H_{c2}/\rho_n c^2$. Here, $\Phi_0$ is the magnetic quantum
flux, $c$ is the speed of light, $\rho_n$ is the resistivity of the normal
phase and $H_{c2}$ is the upper critical field. The first term on the right
hand side of Eq.~(\ref{eq:vf}) accounts for the vortex-vortex interaction and
it is given by
\begin{equation}
\vec{J}(\vec{r})\equiv\Phi_0^2/(8\pi\lambda^3)
K_1(|\vec{r}|/\lambda)\hat{r}~,
\end{equation}
where the function $K_1$ is a Bessel function decaying exponentially for
$|\vec{r}| > \lambda$, and $\lambda$ is the London penetration length
\cite{degennes}. The interaction is cut off at a distance $6\lambda$ to
improve computational efficiency. The second term on the right hand side of
Eq.~(\ref{eq:vf}) describes the interaction between pinning centers, modeled
as localized traps, and flux lines. Here, $\vec{G}$ is the force due to a
pinning center located at $\vec{R}_p$, $l$ is the range of the wells
(typically $l \ll \lambda$), and $p = 1, ..., N_p$ ($N_p$ is the total number
of pinning centers). For the pinning force, we use the following expression:
$\vec{G}(\vec{x})=-f_0\vec{x}(|\vec{x}|-1)^2$, for $|\vec{x}|<1$ and zero
otherwise. In the present simulations we restrict ourselves to the case $T=0$
(see Ref.~\cite{MON-00} for the implementation of thermal noise in MD
simulations).

As we discussed in the introduction, we intend to study the effect of
different boundary conditions on the flux penetration. We start
with an empty system and concentrate all the vortices in a small strip at the
boundary. Due to mutual repulsion, the vortices will be pushed inside the
material, forming a flux front. There are several ways to implement the
boundary conditions, corresponding to different experimental
situations. Here, we will consider the following boundary conditions
\cite{dorog,gilc}:
\begin{itemize}
\item{(A)} Constant total number of vortices. Experimentally this corresponds 
to an external control of the magnetic flux.
\item{(B)} Constant vortex concentration at the boundary. This case
corresponds to an external control of the magnetic field.
\item{(C)} Total vortex number increasing at constant rate. This 
represents an external control of the flux rate.
\item{(D)} Boundary concentration increasing at constant rate,
corresponding to a constant field rate.
\end{itemize}
As a word of caution, one should notice that boundary conditions can be more
complicated in reality, due to complex surface barriers that oppose flux
penetration. These are not considered here and the only surface barrier
is provided by already entered flux lines.

\section{MD simulations}

We perform MD simulations based on Eq.~(\ref{eq:vf}) and analyze the
flux front propagation for different values of the pinning strength
$f_0$.  We typically use up to $N_p=800~000$ Poisson distributed
pinning centers of width $l=\lambda/2$ in a system of size $(L_x=800
\lambda , L_y=100\lambda$), corresponding to a density of $n_0=
10/\lambda^2$.  The number $N$ of flux lines depends essentially on
the boundary condition adopted in the simulation. The injection of
magnetic flux into the sample is implemented as in Ref.~\cite{zms},
concentrating at the beginning of the simulation all the flux lines in
a small strip $L'\ll \lambda$, parallel to the $y$ direction, and
imposing periodic boundary conditions in both directions. The front
position is taken as the $x$ coordinate of the most advanced particle
in the system at different times.

The case (A), corresponding to a constant vortex number, was studied in
detail in Ref.~\cite{zms}, where we showed that the front position $x_p$
grows initially with time as $t^{1/3}$ for small times. Eventually the front
position slows down and saturates to a value $\xi_p$ which increases as the
strength of the pinning centers $f_0$ is decreased. 
In particular, the front pinning length $\xi_p$ was found to scale 
as $ f_0^{-1/2}$ \cite{zms}. Here, we compare this behavior with that 
of cases (B-D).

\begin{figure}  
\epsfig{file=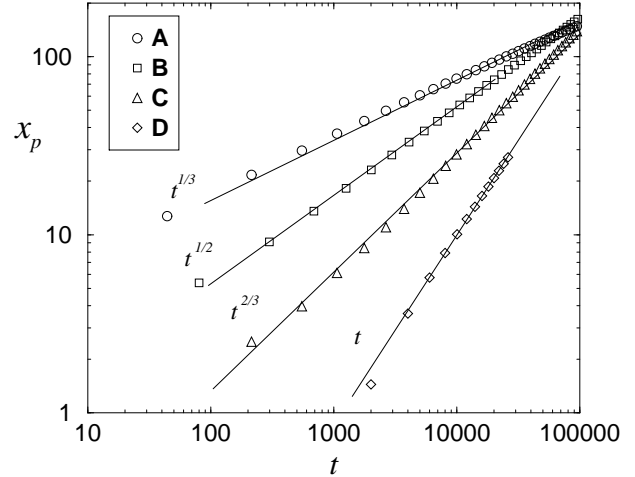,width=8cm} 
\caption{The average position of the front plotted as a function of time.
The data has been obtained from MD simulations with different boundary
conditions (for a definition see text) in a clean system ($f_0=0$). The
curves increase as $t^{\alpha}$, where $\alpha$ depends on the particular
boundary condition imposed.}
\label{fig:1}
\end{figure}

\begin{figure}
\epsfig{file=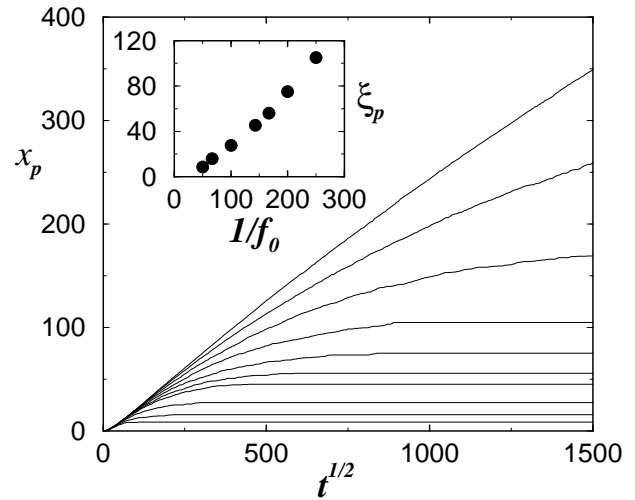,width=8cm}
\caption{The average position of the front, obtained from MD simulations
for case (B), plotted as a function of time. The curves increase as
$t^{1/2}$ and saturates at long times to a value depending on
$f_0$. From top to bottom, the curves correspond to $f_0=0.001$,
$0.002$, $0.003$, $0.004$, $0.005$, $0.006$, $0.007$, $0.01$, $0.015$,
and $0.02$. In the inset, we show that the pinning length scales as
$f_0^{-1}$.}
\label{fig:2} 
\end{figure}

We first consider the case $f_0=0$, corresponding to a clean
superconductor (i.e., without defects) in order to clearly identify
the front penetration law in the initial regime. In Fig.~\ref{fig:1}
we show that the front advances as a power law with an exponent that
depends on the boundary condition.  In particular, we find $x_p \sim
t^{1/2}$ for case (B), $x_p \sim t^{2/3}$ for case (C) and $x_p \sim
t$ for case (D). The presence of disorder can affect all these
behaviors in a different way, depending on the imposed boundary
condition.  The case (B) is quite similar to the case (A) studied in
Ref.~\cite{zms}: after an initial transient, the front gets pinned and
the pinning length scales with the pinning strength (see
Fig.~\ref{fig:2}). In particular, the pinning length scales roughly as
$x_p \sim 1/f_0$, as shown in the inset of Fig.~\ref{fig:2}. For case
(C), the front never gets pinned. The effect of disorder is only to
slow down the dynamics (Fig.~\ref{fig:3}). A similar behavior is found
for case (D).

\section{Non-linear diffusion}

In Ref.~\cite{zms}, we have shown that the front propagation can be described
by coarse graining the system and obtaining a disordered non-linear diffusion
equation. The equation that rules the evolution of the local vortex density
$\rho(\vec{r},t)$ is
\begin{equation}
\Gamma\frac{\partial \rho}{\partial t}=
\vec{\nabla}(a\rho\vec{\nabla}\rho-\rho\vec{F}_c)
+k_BT \nabla^2\rho~,
\label{eq:fin}
\end{equation}
where $a=\Phi_0^2/4$ and $F_c$ is a random {\it friction} force due to the
pinning centers, with a typical value scaling as $F_c\sim f_0\sqrt{n_0}$.

For $T=0$ and $f_0=0$, Eq.~(\ref{eq:fin}) can be solved exactly using scaling
methods \cite{dorog,gilc}. In particular the density profiles obey the
equation
\begin{equation}
\rho(x,y,t)=t^{-\alpha} h( x/t^{\beta})~,
\end{equation}
where $\alpha$ and $\beta$ depend on the boundary conditions and satisfy
$\alpha+2\beta=1$. For the cases considered : A) $\alpha=1/3$, $\beta=1/3$;
B) $\alpha=0$, $\beta=1/2$; C) $\alpha=-1/3$, $\beta=2/3$; and D) $\alpha=-1$,
$\beta=1$.  These results are in perfect agreement with the numerical
simulations reported in Fig.~\ref{fig:1}, since the exponent $\beta$
describes the dynamics of the front position.

\begin{figure}  
\epsfig{file=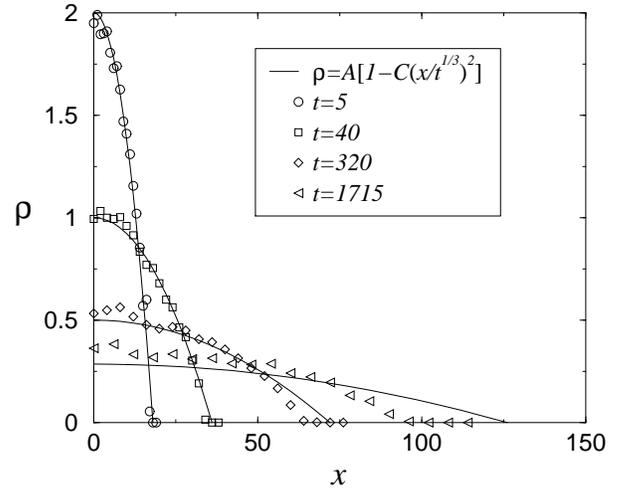,width=8cm} 
\caption{The density profile at different times $t$ obtained from MD
simulations for boundary condition (A) in the presence of disorder. At the
beginning the profile follows the solution of the non-linear diffusion
equation in the disorder free case and eventually deforms due to the action
of pinning.}
\label{fig:4} 
\end{figure}

The function $h(u)$ also depends on the boundary condition and for the
case (A) is given by $h(u)=(1-u^2)/6$ for $u<1$ and vanishes for
$u\geq 1$. The other cases are reported in Refs.~\cite{dorog,gilc}. 
We check by MD numerical simulations that, in the presence of disorder,
the density profiles are described by the non-linear diffusion
equation (\ref{eq:fin}). In Fig.~\ref{fig:4} we show that the profiles
follow the $f_0=0$ solution and then deform when pinning starts to
dominate.

\section{Front dynamics}

In order to understand in a simple way the effect of disorder on the front
propagation for different boundary conditions, we can write an equation for
the average position of the front. The approach is very similar in spirit to
what is done for the imbibition of porous media \cite{dube}. As discussed
above, the front is driven by the density gradient against the pinning
landscape. The density gradient can be estimated simply as $\nabla \rho \sim
\rho(0,t)/x_p$, where $x_p$ is the front position and $\rho(0,t)$ is the
boundary density. The typical pinning force can simply be taken as
$f_0\sqrt{n_0}$.  Collecting these two contribution we write
\begin{equation}
\Gamma dx_p/dt = a\rho(0,t)/x_p -f_0\sqrt{n_0}~.
\label{eq:bc}
\end{equation}
In order to close the problem we have to specify the behavior of the
boundary density, which clearly will  depend on the particular boundary
condition chosen. Let us consider the various cases:

\noindent(A) When the total number of vortices is conserved, the boundary
density decreases as the front advances. This can be explained by noting that
the total number of vortices can be roughly estimated as $M=\rho(0,t) x_p$,
so that $\rho(0,t)=M/x_p$. Inserting this into Eq.~(\ref{eq:bc}) we obtain
\begin{equation}
dx_p/dt = 1/x_p^2 -g,
\label{eq:bc-a}
\end{equation}
where $g_A=f_0\sqrt{n_0}/a$ and time is expressed in units of
$\Gamma/(Ma)$. Eq.~(\ref{eq:bc-a}) admits an implicit solution as
\begin{equation}
(g/M)t=\mbox{arctanh}(\sqrt{g/M}x_p)/\sqrt{g/M}-x_p~,
\end{equation}
which can be expanded for $x_p \ll 1/\sqrt{g/M}$ to give $x_p\sim t^{1/3}$
(short times), and for $x_p \simeq 1/\sqrt{g/M}$ (long times) yielding
$x_p\simeq \sqrt{M/g}(1-2\exp(-2(g/M)^{3/2} t))$, corresponding to a pinned
front. This behavior is in agreement with the scaling found in numerical
simulations (see Ref.~\cite{zms}).

\noindent(B) The case of a constant vortex density at the boundary is similar
to case (A). The boundary condition is simply $\rho(0,t)=\rho_0$, and
Eq.~(\ref{eq:bc}) becomes
\begin{equation}
dx_p/dt = \rho_0/x_p-g,
\label{eq:bc-b}
\end{equation}
where time is now expressed in units of $\Gamma/(\rho_0a)$.
As in the previous case, Eq.~(\ref{eq:bc-b})
can not be solved explicitly but from the implicit solution it is possible to
obtain the asymptotic behavior: $x_p \sim t^{1/2}$ at short times and
$x_p\simeq \rho_0/g(1-\exp(-(g/\rho_0)t))$ at long times, in agreement with the
results presented in Fig.~\ref{fig:2}.

\begin{figure}  
\epsfig{file=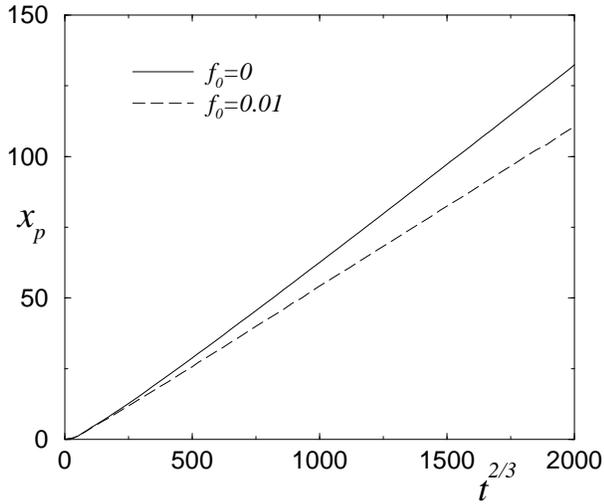,width=8cm} 
\caption{The average position of the front for case (C) obtained from MD 
simulations. The solid line corresponds to a clean system and the
dashed line to a disordered system. The front moves as $x_p \sim
At^{2/3}$, and $A$ is reduced in the presence of disorder.}
\label{fig:3} 
\end{figure}

\noindent(C) This case is similar to case (A), with the difference that the
total number of vortices increases with time (i.e. $M=ht$).  Due to this the
front is never pinned. In absence of pinning we recover the $t^{2/3}$
behavior observed in MD simulations.  We can not find an analytical solution
of the equation in this case and resort to numerical integration. The results
presented in Fig.~\ref{fig:3} indicate that the front asymptotically grows as
$x_p=C+At^{2/3}$, where $A$ is the solution of the equation $A^3=3/2(h-gA^2)$. 
Thus for low pinning we expect that the coefficient $A$ decreases linearly 
with the pinning strength. This result is in agreement with the MD simulation 
(see Fig.~\ref{fig:3}).

\noindent(D) As in case (C), the front is not pinned by the disorder, which
has the only effect of reducing the front velocity. One can compute the
asymptotic velocity imposing $x_p=Vt$ and inserting this expression in 
Eq.~(\ref{eq:bc}), that now reads
\begin{equation}
dx_p/dt = ht/x_p-g,
\label{eq:bc-d}
\end{equation}
where $h$ is the rate of increase of the boundary density. 
Solving for $V$ one obtains $V=(\sqrt{g^2+4h}-g)/2$.

\section{discussion and perspectives}

In this paper we have analyzed the effect of different boundary
conditions on the flux penetration in disordered type II
superconductors. We have conducted a series of MD simulations of
interacting vortices and interpreted the results in terms of a
non-linear diffusion equation. In the limit of no disorder, the
equation has been solved in Refs.~\cite{dorog,gilc} yielding solutions
for the front propagation and the density profiles. Here we have shown
that these results are in perfect agreement with MD simulations. Moreover, 
we have found that the presence of pinning centers affects
the behavior of the system and, depending on the boundary conditions,
the front is either pinned or simply slowed down. To clarify these
effects, we have introduced a simple equation of motion for the front
position, in the same spirit of Washburn approach to imbibition
\cite{dube}. Despite its simplicity, the equation captures the
essential features of the front dynamics.

An important question that still remains to be addressed is the relevance of
these results for experiments. We can associate each one of the boundary
condition studied here to different experimental conditions, corresponding to
the way the field is applied to the sample. In many cases, the presence of
surface barriers for flux penetration could in principle modify the scaling
behavior discussed here. We believe, however, that away from the surface this
effect should not be important and the scaling would be recovered. In
addition, thermal and quantum creep effects in general may lead
to a slowly moving front, even when pinning is expected. It is possible to
account for these effects in this framework considering the diffusion term in
Eq.~(\ref{eq:fin}) or adding a random noise term to the front propagation
equation. In conclusion, we expect that our approach together with a
systematic series of magneto-optical measurements performed under different
magnetic field controls (e.g., see Refs.~\cite{rough,fractal,fractal2}, will
represent significant steps for the understanding of the flux penetration
phenomenon in disordered superconductors.

\section{acknowledgement}
This work has been supported by CNPq and FUNCAP. S. Z. wishes to thank for
hospitality at the physics department of UFC where this work has been
completed.

\end{multicols}

\begin{thebibliography}{99}
\bibitem{Blatter}
	G.~Blatter, M.V.~Feige{l'}man, V.B.~Geshkenbein, A.I.~Larkin,
	and V.M.~Vinokur, Rev.~Mod.~Phys.~{\bf 66}, 1125 (1994).
\bibitem{bean}
	C. P. Bean, Rev. Mod. Phys. {\bf 36}, 31 (1964).
\bibitem{kim}
        Y. B. Kim, C. F. Hempstead and A. R. Strnad,
        Phys. Rev. {\bf 129}, 528 (1963).
\bibitem{avalanches}
        S. Field, J. Witt, F. Nori, and X. Ling,
        Phys. Rev. Lett. {\bf 74}, 1206 (1995);
        C. M. Argenter, Phys. Rev. B {\bf 58},
        1438 (1998);
        K. Behnia, C. Capan, D. Mailly and B. Etienne,
        Phys. Rev. B {\bf 61}, R3815 (2000).
\bibitem{rough}
         R. Surdeanu, R. J. Wijngaarden, E. Visser, J. M. Huijbregtse, 
	J. H. Rector, B. Dam, and R. Griessen 
        Phys. Rev. Lett. {\bf 83}, 2054 (1999).
\bibitem{fractal}
        R. Surdeanu, R. J. Wijngaarden, B. Dam, J. Rector, R. Griessen,
	 C. Rossel, Z. F. Ren, and J. H. Wang, Phys. Rev. B {\bf 58},
        12467 (1998). 
\bibitem{fractal2}
	S. S. James, S. B. Field, J. Seigel and H. Shtrikman,
	Physica C {\bf 332}, 445 (2000).
\bibitem{Brass}
	H. J.~Jensen, A.~Brass and A.J.~Berlinsky, 
	Phys.~Rev.~Lett.~{\bf 60}, 1676 (1988).

\bibitem{nori2}
C. Reichhardt, J. Groth, C. J. Olson, S. Field, and F. Nori. 
     Phys. Rev. B {\bf  52}, 10441 (1995); {\bf 53}, R8898 (1996).

\bibitem{pla}
O. Pla, N. K. Wilkin and H. J. Jensen,
Europhys. Lett. {\bf 33}, 297 (1996).

\bibitem{nori4}
C. J.~Olson, C.~Reichhardt, and F.~Nori, 
Phys.~Rev.~B {\bf 56}, 6175 (1997);
Phys.~Rev.~Lett. {\bf 80}, 2197
(1998). 


\bibitem{zms}
S. Zapperi, A. A. Moreira, J. S. Andrade Jr., Phys. Rev. Lett. 
{\bf 86}, 3622 (2001).

\bibitem{dorog}
V. V. Bryskin and S. N. Dorogotsev,
JEPT {\bf 77}, 791 (1993); Physica C {\bf 215}, 173 (1993).

\bibitem{gilc}
J. Gilchrist and C. J. van der Beek,
Physica C {\bf 231}, 147 (1994).


\bibitem{degennes}
        P.-G. de Gennes, {\em Superconductivity of metals
        and alloys} (Benjamin, New York, 1966).

\bibitem{MON-00}
	D. Monier and L. Fruchter,
	Eur. Phys. J. B {\bf 17}, 201 (2000).

\bibitem{dube}
	E. Washburn, Phys. Rev. {\bf 17}, 273 (1921).
	For a review see
	 M. Dube, M. Rost and M. Alava,
	Eur. Phys. J. B {\bf 15}, 691 (2000).


\end{thebibliography}
\end{document}